\PassOptionsToPackage{svgnames}{xcolor}
\documentclass{article}
\usepackage{authblk}
\usepackage[scaled=.89]{helvet}
\usepackage{graphicx} 
\usepackage{booktabs}
\usepackage{soul}
\usepackage{array}
\usepackage{enumitem}
\usepackage{todonotes}
\usepackage[svgnames]{xcolor}
\usepackage{amsmath}
\usepackage{comment}
\usepackage{placeins}
\usepackage[frozencache]{minted}
\definecolor{MediumSlateBlue}{rgb}{0.48, 0.41, 0.93}
\definecolor{BurntOrange}{RGB}{190,81,3}
\usepackage{listings}
\usepackage[colorlinks,linkcolor=MediumSlateBlue,citecolor=MediumSlateBlue,urlcolor=MediumSlateBlue]{hyperref}
\usepackage{cleveref}
\usepackage{marginnote}
\usepackage{xspace}

\newcommand{\fb}[1]{\todo[color=green!30]{\footnotesize #1}}
\newcommand{\fbL}[2][]{{%
 \let\marginpar\marginnote
 \reversemarginpar
 \todo[#1,color=green!30]{\footnotesize #2}}}

\newcommand{\cplusplus}{C\texttt{++}\xspace}
\newcommand{\cppstd}[1]{C\texttt{++}#1\xspace}

\newcommand{\ignore}[1]{}

\newcolumntype{L}[1]{>{\raggedright\let\newline\\\arraybackslash\hspace{0pt}}m{#1}}
\newcolumntype{C}[1]{>{\centering\let\newline\\\arraybackslash\hspace{0pt}}m{#1}}
\newcolumntype{R}[1]{>{\raggedleft\let\newline\\\arraybackslash\hspace{0pt}}m{#1}}

\title{
Interface for Sparse Linear Algebra Operations}

\author[1,2]{Hartwig Anzt}
\author[2]{Ahmad Abdelfattah}
\author[3]{Willow Ahrens}
\author[4]{Chris Armstrong}
\author[5]{Benjamin Brock}
\author[6,14]{Aydin Buluc}
\author[7]{Federico Busato}
\author[8]{Terry Cojean}
\author[9]{Tim Davis}
\author[6]{Jim Demmel}
\author[6]{Grace Dinh}
\author[10]{David Gardener}
\author[11]{Jan Fiala} 
\author[2]{Mark Gates}
\author[7]{Azzam Haider}
\author[12]{Toshiyuki Imamura}
\author[13]{Pedro Valero Lara}
\author[14]{Sherry Li}
\author[3,2]{Piotr Luszczek}
\author[2]{Max Melichenko}
\author[11]{Yvan Mokwinski}
\author[15]{Jose Moreira}
\author[16]{Riley Murray}
\author[17]{Spencer Patty}
\author[13]{Slaven Peles}
\author[1]{Tobias Ribizel}
\author[11]{Jason Riedy}
\author[16]{Siva Rajamanickam}
\author[13]{Piyush Sao}
\author[11]{Manu Shantharam}
\author[13]{Keita Teranishi}
\author[7]{Stan Tomov}
\author[1]{Yu-Hsiang Tsai}
\author[18]{Heiko Weichelt}
\affil[1]{TU Munich, Germany}
\affil[2]{University of Tennessee, USA}
\affil[3]{Massachusetts Institute of Technology (MIT), USA}
\affil[4]{arm, UK}
\affil[5]{Intel Labs, USA}
\affil[6]{University of California, Berkeley, USA}
\affil[7]{NVIDIA, USA}
\affil[8]{EVIDEN, France}
\affil[9]{Texas A\&M University, USA}
\affil[10]{Lawrence Livermore National Labs, USA}
\affil[11]{AMD, USA}
\affil[12]{RIKEN Center for Computational Science, Japan}
\affil[13]{Oak Ridge National Lab, USA}
\affil[14]{Lawrence Berkeley National Labs, USA}
\affil[15]{IBM, USA}
\affil[16]{Sandia National Labs, USA}
\affil[17]{Intel Corporation, USA}
\affil[18]{The Math Works Inc, USA}

\date{\today}

\begin{document}

\maketitle

\section*{Abstract}
The standardization of an interface for dense linear algebra operations in the BLAS standard has enabled interoperability between different linear algebra libraries, thereby boosting the success of scientific computing, in particular in scientific HPC. In particular, from this point on, applications could interface highly optimized implementations through a standardized API. 
Despite numerous efforts in the past, the community has not yet agreed on a standardization for sparse linear algebra operations due to numerous reasons. One is the fact that sparse linear algebra objects allow for many different storage formats, and different hardware may favor different storage formats. This makes the definition of a FORTRAN-style all-circumventing interface extremely challenging. Another reason is that opposed to dense linear algebra functionality, in sparse linear algebra, the size of the sparse data structure for the operation result is not always known prior to the information. Furthermore, as opposed to the standardization effort for dense linear algebra, we are late in the technology readiness cycle, and many production-ready software libraries using sparse linear algebra routines have implemented and committed to their own sparse BLAS interface.
At the same time, there exists a demand for standardization that would improve interoperability, and sustainability, and allow for easier integration of building blocks.
In an inclusive, cross-institutional effort involving numerous academic institutions, US National Labs, and industry, we spent two years designing a hardware-portable interface for basic sparse linear algebra functionality that serves the user needs and is compatible with the different interfaces currently used by different vendors. In this paper, we present a C++ API for sparse linear algebra functionality, discuss the design choices, and detail how software developers preserve a lot of freedom in terms of how to implement functionality behind this API.
\newpage
\tableofcontents

\section{Introduction and Motivation}

Despite the fact that Sparse matrix computations are at the heart of many science and engineering applications the Sparse BLAS standard~\cite{10.1145/567806.567810} proposed decades ago has not gained enough traction in the community to become a standard allowing for portability across different hardware architectures. A reason for this may be that it was focused on CPU architectures, not considering the plethora of optimization options relevant to today's accelerator architectures, such as GPUs. 

In the meantime, many vendors already provide support for sparse matrix computations in proprietary or open-source libraries, but due to diverging architectural constraints and new functionality required, these libraries have different execution models, APIs, and formats supported. At the same time, the well-recognized and successful BLAS standard for dense linear algebra has been modernized by the \cplusplus committee when integrating the functionality into \texttt{std::linalg} when integrating the functionality into the \cplusplus-26 standard library.

Learning from these previous and ongoing efforts and building on top of the \cplusplus committee's recent work, we propose a new \cplusplus-based Sparse BLAS standard. Our main objectives are to enable portability across vendors, and to provide high-performance, flexible and extensible interfaces for sparse linear algebra functionality. Our API is designed to run on accelerators as well as traditional CPUs alike. This is especially important given the pervasiveness of accelerators in computing and the challenges in achieving high performance from such hardware, which further increases the value of Sparse BLAS.

We recognize that linear algebra is a fundamental topic beyond classical mathematical disciplines and physical sciences. Over the past decade, sparsity has emerged as an important element of computation in AI. We argue that a Sparse BLAS standard would provide a useful set of tools for modern AI frameworks, potentially expanding to standardize new functionality such as low- and mixed-precision computation and batch processing, and easing integration with high-level deep learning frameworks.

While we may include more high-level routines in the future, such as solvers, we initially restrict ourselves to basic functionality and also retain the familiar name BLAS (Basic Linear Algebra Subroutines). Given the breadth and scope of Sparse functionality (numerous matrix formats, complex algorithms, ...), we expect the standardization effort to be an iterative process. We focus in the current iteration on functionalities for which we found there is sufficient interest and which are achievable for a first version of this standard. Therefore, the list of features, kernels, and options may expand in future. The current API design proposal is the fruit of an open, collegial collaboration between industry and university experts. To suggest functionality for inclusion, or to get involved in the project, feel free to contact the coordinator of the effort as identified in the author list.

Concerning interfaces, we currently focus on the \cplusplus interface as the key building block, but plan to also suggest interfaces for C and Fortran.  We acknowledge the use of high-level languages such as Python and Julia in the HPC community and plan to provide interfaces also to high-level programming languages in a later iteration.

\section{Related Efforts}

\subsection{The (dense) BLAS standard}
The origins of the BLAS standard can be
traced back to 1973 when Hanson, Krogh, and Lawson wrote an article in the
SIGNUM Newsletter (Vol. 8, no. 4, p. 16) describes the advantages of adopting
a set of basic routines for problems in linear algebra. This led to the
development of the original BLAS~\cite{Lawson1979l1blas}, which indeed turned out to be
advantageous and very successful. It was adopted as a standard and used in a
wide range of numerical software, including LINPACK~\cite{Dongara:1979:linpack}. An extended, Level 2 BLAS, was proposed for matrix-vector operations~\cite{dongarra1988a,dongarra1988b}.
At the time, these two levels were appropriate for vector processors and
vector register chaining, respectively.
Later in the 90s with the appearance of cache-based machines, a strong need to restructure existing algorithms to benefit from reusing data from cache emerged.
The idea was to express algorithms in terms of blocks that fit into cache memory,
decreasing data movement by providing a surface-to-volume effect for
the ratio of $O(n^2)$ data movement to $O(n^3)$ operations.
Therefore, the Level 3 BLAS were proposed~\cite{dongarra1990a,dongarra1990b},
covering matrix-matrix operations, and LINPACK was redesigned
into LAPACK~\cite{anderson1999} to use the new Level 3 BLAS where possible.
With multi-core and GPU-accelerated architectures, the BLAS have continued to
be a valuable paradigm, including in parallel task-based libraries such
as PLASMA~\cite{agullo2009numerical}.

Several \cplusplus interfaces have been proposed for the BLAS,
ranging from lightweight wrappers~\cite{gates2017cpp} to
expression templates~\cite{siek1998matrix,boost_ublas,eigen} that replace
function calls with arithmetic expressions (\verb+C = A*B+).
Recently, the \verb+std::linalg+ interface~\cite{hoemmen2023linalg}
was adopted as part of the
\cppstd{26} standard, and has served as an inspiration
for this Sparse BLAS proposal.
\texttt{linalg} is a templated, free-function interface that leverages
\texttt{mdspan} as an
abstraction for dense matrices and tensors.
Matrix scaling and (conjugate) transposition are provided by lightweight
views of the matrix, which we have followed here.
In a break from traditional
BLAS, functions use plain English names (\verb+multiply+) instead
of expert-focused acronyms (\verb+gemm+). Execution policies provide
a mechanism for parallelization and GPU acceleration.




\subsection{The GraphBLAS standard}

Many graph algorithms can be expressed in the language of linear algebra, often with no loss of performance. Expressing graph algorithms using linear algebra exposes natural parallelism, allowing users to capitalize on optimized sparse linear algebra routines for multi-threaded, GPU, and multi-node execution. This, together with the fact that most large graphs are sparse, positions sparse matrix algebra as a viable abstraction for graph computations~\cite{kepner2011graph, bulucc2011combinatorial}. 

The GraphBLAS API, which provides generalized sparse linear algebra primitives for implementing graph algorithms, capitalizes on the duality between sparse matrices and graphs~\cite{brock2021graphblas}. The GraphBLAS interface introduces several generalizations to traditional linear algebra, including semirings, which allow operations to be performed using different sets of operators than the plus and times used in traditional arithmetic semiring.  Different semirings are necessary for different algorithms: for example, an implementation of single-source shortest path (SSSP) might use a matrix-vector multiply with the plus-min semiring, using plus to add the edge length to the current shortest path, followed by min to update the path to the destination only if a shorter path has been found.  Semirings allow the same algorithm implementations to be used with many different graph algorithms.  GraphBLAS also provides a few other features necessary for graph computation, such as masked operations, permutations, and element-wise operations.

The GraphBLAS C API Specification~\cite{brock2021graphblas} defines a
set of opaque containers such as matrices and vectors along with
generalized linear algebra operations in C.  SuiteSparse provides a
high-performance, multi-threaded implementation of the GraphBLAS C API,
along with some support for GPU
execution~\cite{davis2019algorithm,davis2023algorithm}. There are also
several GraphBLAS-like libraries in other programming
libraries~\cite{erik_welch_2024_10631255, yzelman2020c++,
yang2022graphblast}.  The GraphBLAS \cplusplus Specification provides
a native \cplusplus interface, providing better support for
user-defined types through templates as well as taking advantage of
\cplusplus concepts and views to simplify the
interface~\cite{9150467,graphblascppspec}.

While GraphBLAS provides much of the functionality necessary needed in a Sparse
BLAS interface, there are several issues that make it suboptimal for adoption
as a Sparse BLAS standard:

\begin{enumerate}
\item While GraphBLAS' opaque objects
provide a clean separation between what data is owned by the user and what data
is owned by the implementation, this separation generally requires data to be
copied in and out of GraphBLAS via its import/export API.  For users who only
want to use a library to execute a few operations, this presents considerable
overhead.  This issue can be resolved with the pack/unpack methods in
SuiteSparse:GraphBLAS, which takes $O(1)$ time and does not allocate memory. 
These methods require the library and GraphBLAS to share the same memory space
and are extensions to GraphBLAS that are not in the GraphBLAS C API
Specification, however.

\item GraphBLAS provides some complex functionality, such as
masked operations and extract and assign routines, that are not needed by
Sparse BLAS, and that vendors may be unlikely to implement.
The GraphBLAS C API allows an unimplemented method to return an error code to
indicate that it is not implemented, so this issue could be resolved by
creating a well-defined Sparse BLAS subset of GraphBLAS.

\item Sparse
BLAS has a few key routines, such as triangular solve, that are not needed or
included in GraphBLAS.  

\item
The GraphBLAS \cplusplus API does not currently provide explicit support
for accelerators or explicitly control how and where code should be
executed, although the \cplusplus specification does allow
implementations to support accelerators through \cplusplus execution
policies~\cite{9150467,graphblascppspec}. Parallel and accelerated
execution should be a first-class consideration for a SparseBLAS
specification.

\item
Finally, there are currently no vendor-provided implementations of GraphBLAS, with vendors deferring to support open-source implementations where necessary due to the breadth of the API.  This Sparse BLAS effort includes math library developers from many vendors (Intel, NVIDIA, AMD, Arm), and we anticipate working with vendors to ensure the interface is suitable for them to produce high-performance implementations.

\end{enumerate}

Nevertheless, we anticipate that the proposed standard and GraphBLAS will overlap for a subset of functionalities, and the Sparse BLAS API will be capable of efficiently expressing them. Several GraphBLAS innovations we may explore utilizing in this iteration of the Sparse BLAS standard include: (1) operations between two sparse objects, (2) masking, and (3) non-blocking or asynchronous execution.

\subsection{Existing Sparse BLAS standards}

The first mention of a Sparse BLAS standard was in 1985, with Dodson et
al.~\cite{10.1145/1057935.1057938} calling for Sparse extensions to the BLAS standard, recognizing the growing importance of sparse
solvers and preconditioners in research and industry. At the time, the proposed
extensions primarily focused on level-1 operations (i.e., vector-vector
operations) and the use of compressed vector storage format. Many important
functions such as scatter, gather, and dot products are proposed with interfaces in
this short paper. The original SIGNUM newsletter paper was extended in
1991~\cite{10.1145/108556.108577} but the focus on vector operations was kept.

Recognizing the lack of a matrix-focused Sparse BLAS proposal, in 1997, Duff et
al.~\cite{10.1145/275323.275327} proposed a level-3 BLAS interface design (i.e.,
matrix-matrix operations). In this paper, they propose 1) a sparse-matrix
dense-matrix product, 2) upper and lower triangular solves, 3) utility functions
to convert between matrices and scale matrices, and 4) routines to permute the
rows and columns of matrices. They also integrate workspace management into the
interface with appropriate error handling due to workspace allocations being a
necessity in several sparse-matrix operations. Thanks to Fortran 90, they also
show how to implement generic interfaces (e.g., called multiply for matrix
products, with a \texttt{spmat} type representing sparse matrices) that is simpler
to use compared to routines with letter codes and long argument lists.

In 2002, Duff et al.~\cite{10.1145/567806.567810} grouped the previous
proposals into a new standardization proposal of operations on sparse matrices
and vectors. This includes operations equivalent to the dense BLAS level 1, level 2, and level 3 operations, including sparse triangular solves, as well as operations for the creation of sparse matrices and inserting nonzero
elements into them. The proposed standard recognizes that almost
every application area has a different way of storing and accessing the nonzero
entries of the sparse matrix, making it difficult to design software applicable to
all areas~\cite{10.1145/567806.567810}. In response, the proposed interface takes
as input not the matrix entries themselves, but rather a pointer, or a handle to
a generic representation of a previously created sparse matrix object. This
allows algorithms to be coded using Sparse BLAS primitives without disclosing
the underlying matrix storage format. The authors do suggest a Fortran-style
interface for accessing Sparse BLAS functionality from C and Fortran.  The
proposed 2002 SparseBLAS standard does not support operations between two sparse
objects (matrices or vectors) due to perceived complexity, which we plan to
address in this iteration.

In parallel with the previous efforts, several libraries with Sparse BLAS
functionality were developed. One such library that stands out is PSBLAS from
Filippone et al.~\cite{10.1145/365723.365732}. The scope and breadth of the
library make it akin to a Sparse standard proposal, as most operations are
supported and careful attention is given to interface design. They also
reuse the work of Duff et al.~\cite{10.1145/275323.275327}, e.g. use the same
matrix storage type and rely on the serial Sparse BLAS of the previous
proposals. Their focus is on proposing a parallel (node-distributed)
implementation of the algorithms and implementing the library using
multiple layers, with base functionality implemented in Fortran 77 and
higher-level interfaces in Fortran 90. In 2012, Filippone et
al.~\cite{10.1145/2331130.2331131} proposed an Object-Oriented, Fortran 2003,
interface thanks to recent programming language innovation that greatly
simplifies Sparse BLAS usage and standardization. This design enables the support
of multiple formats through class inheritance while maintaining an abstract
matrix handle type. This handle type allows for conversions and other optimizations while
enabling simpler interfaces. Notably, the design also considers move versus copy
semantics for matrix conversions and user data.

We learn from and base our standard proposal upon the previous efforts to provide an
increasingly simple-to-use yet expressive Sparse BLAS set of functionality. At
the same time, we extend the scope by, e.g., introducing an interface for
the more complex sparse-matrix times sparse-matrix product deemed
complicated in the previous iterations. The hardware evolution, namely GPUs,
requires us to also introduce a new set of functionality, user-provided allocators, as memory
management can be critical for these complex operations on such complex
hardware. 
Furthermore, we acknowledge in this proposal that for repeated application of a sparse operator, the initial investment in optimization -- analyzing sparsity patterns and characteristics, and fine-tuning format and execution policies -- can quickly be amortized by faster execution. We thus propose interfaces also allowing for an optimization phase prior to repeated execution of functionality.

\section{Functionality Scope of the Sparse BLAS}

The goal of this effort is to define a common understanding of how to access a set of standard data structures representing sparse linear algebra objects and a common interface for basic functionality that operates on these objects.
To reduce complexity, we have omitted advanced functionalities that operate on sparse objects such as iterative solvers, eigensolvers, and preconditioners. Even so, the Sparse BLAS serves as an important foundation for building high-level applications, facilitating integration between advanced algorithms and low-level building blocks.

Our scope entails \textit{sparse storage format standardization} as well as \textit{conversion between formats}, the \textit{multiplication and addition of matrices} (including operations involving both sparse and dense matrices), \textit{transposition and conjugation} of sparse matrices, \textit{a solver for sparse triangular linear systems}, \textit{scaling} of sparse matrices, \textit{norm} calculations on sparse matrices, \textit{element-wise addition and multiplication}, and \textit{masking operations} that extract a subset of the nonzero entries of a sparse matrix as in the sampled dense times dense matrix multiplication (SDDMM) kernel common in machine learning frameworks. A list of functionality falling into this scope is given in \Cref{tab:scope}. For convenience, we also summarize the API type for each functionality as detailed in \Cref{sec:api-design}. Multi-stage APIs produce a Sparse Matrix as an output, whereas single-stage APIs produce a dense object as an output.

\begin{table}[h]
    \centering
    \footnotesize
    \begin{tabular}{lll}
    \toprule
    \textbf{Operation} & \textbf{Notation} & \textbf{API Type}\\ 
    \midrule
       Scaling                                            & $A := \alpha A$ & Single-stage \\
       Transpose and Conjugate Transpose                  & $A := A^T$, $A := A^H$ & Multi-stage\\
       Infinity Matrix Norm                               & $\alpha := \|A\|_\text{inf}$ & Single-stage \\
       Frobenius Matrix Norm                              & $\alpha := \|A\|_F$  & Single-stage \\
       Element-wise Multiplication                        & $C := A~.*B$  & Multi-stage\\
       Sparse Matrix -- Sparse Matrix Addition            & $C := A+B$ & Multi-stage\\
       Sparse Matrix -- Sparse Matrix Multiplication      & $C := A \cdot B + D$ & Multi-stage\\
       Sparse Matrix -- Dense Matrix Multiplication       & $Y^\prime := \alpha A \cdot X + \beta Y$  & Single-stage \\
       Sparse Matrix -- Dense Vector Multiplication       & $y^\prime := \alpha A \cdot x + \beta y$  & Single-stage \\
       Triangular Solve                                   & Solve $A \cdot y = x$ for $y$  & Single-stage \\
       Sparse Matrix Format Conversion                    & $B = \text{sparse}(A)$ & Multi-stage\\
       Predicate Selection                                &  $B = A( \text{predicate} )$ & Multi-stage \\
       Sampled Dense -- Dense Matrix Multiplication       & $C\langle \text{mask} \rangle = A\cdot B$  & Single-stage\\
    \bottomrule
    \end{tabular}
    \caption{List of the Sparse BLAS functionalities. Uppercase letters represent matrices; lowercase represent vectors. Letters from the start of the alphabet are used for sparse matrices/vectors; letters from the end are used for dense matrices/vectors. $\alpha$ is a scalar value. Note that scaling may be applied to any input matrix/vector, and the transpose/conjugate transpose operation may be applied to any input matrix.} 
    \label{tab:scope}
\end{table}

\section{Supported Sparse Matrix Storage Formats}
\label{sec:sparse-formats}

Over the decades, a set of commonly used sparse matrix formats has emerged, providing a versatile set of data structures that allow for productivity and performance. The formats we identified as most common are \textit{Compressed Sparse Row} (CSR)\footnote{Also known as Compressed Row Storage (CRS)}, \textit{Compressed Sparse Column} (CSC), and \textit{Coordinate} (COO).
These formats are well-supported in existing libraries and widely used
for storage and exchange. We will consider these as input formats for
the standard, and each of these formats will be supported through a lightweight, non-owning view class defined by the Sparse BLAS standard. We will also support dense vectors and matrices, including row and column major versions, using \cppstd{23}'s \texttt{mdspan} as a standard view interface for dense vector and matrix formats. The proposed set of user inputs necessary for defining a matrix stored in these formats as well as constructing a new matrix type are given below. It is noteworthy that the API can be extended to accommodate non-standard formats by introducing additional views, removing the need to otherwise modify program logic.

Several additional storage formats have been proposed in the literature throughout the years~\cite{langr2015evaluation}, outlining trade-offs between performance, memory footprint, and creating overhead. These formats can be considered as optimization targets, rather than input/exchange formats. 
Possible future extension formats include CSR(4), CSC(4), BSR / BSC, variable BSR / BSC, Ellpack (including variants ESB/SELL-P/SELL-C-$\sigma$), and diagonal/tridiagonal/pentadiagonal formats. We will not go into specific details about these here; for now, these are considered to be optional extension formats.

Some applications require the structure of sparse matrices only (i.e.,~the pattern of non-zero elements) and not the values themselves. Other applications may use the same value in all nonzero locations, we may call these iso-valued matrices. We do not provide a specific type for structure-only matrices but instead define iso-valued matrices.  We anticipate iso-valued matrix views being created via CSR, CSC, or COO matrix views whose value arrays are replaced by a symbolic ``iso array'' whose values are all the same.

Table \ref{tab:csr_param} outlines the input parameters required to construct a CSR matrix. These input parameters are defined by three important types: \texttt{scalar\_type}, which is the type of the scalar elements stored in the sparse matrix, \texttt{index\_type}, which is the type used to store explicit indices, and \texttt{offset\_type}, which is the type used to store offsets into other arrays.  The \texttt{index\_type} limits the maximum dimensions of a matrix that can be stored, while \texttt{offset\_type} limits the maximum number of nonzeros that can be stored.  This is useful, for example, when large matrices may require \texttt{int64\_t} for \texttt{rowptr} elements in order to avoid integer overflow during calculations involving \texttt{rowptr}.
However, using \texttt{int64\_t} for the large \texttt{colindx} would be unnecessary unless the matrix dimensions are larger than $2^{32}$ and could result in lower performance than using a smaller integer type.
\begin{table}[htbp]
\scriptsize
\begin{center}
     \begin{tabular}{l l l}
         \toprule
         \textbf{Type}      & \textbf{Name} & \textbf{Description} \\
         \midrule
         \texttt{index\_type} & \texttt{nrows} & Number of rows of the matrix \\
         \texttt{index\_type} & \texttt{ncols} & Number of columns of the matrix \\
         \texttt{offset\_type} & \texttt{nnz} & Structural number of non-zeros \\
         \texttt{offset\_type} array & \texttt{rowptr} & Row pointer array (length: \texttt{nrows+1}) \\
         \texttt{index\_type} array & \texttt{colindx} & Column indices array (length: \texttt{nnz}) \\
         \texttt{scalar\_type} array & \texttt{values} & Structural values (length: \texttt{nnz}) \\
         \texttt{base\_type} & \texttt{index\_base} & Base indexing\\
         \bottomrule
     \end{tabular}
\end{center}
     \caption{CSR input parameters}
    \label{tab:csr_param}
\end{table}

The parameter \texttt{nnz} might easily be deduced from the input arrays,
\[\textrm{nnz} = \textrm{rowptr}[\textrm{nrows}] - \textrm{rowptr}[0],\] 
but it can be convenient to store it explicitly in specific scenarios. Therefore, the parameter \texttt{nnz} is part of the sparse data structures.

Parameters for CSC and COO are given in Tables \ref{tab:csc_param} and \ref{tab:coo_param}, respectively.




\begin{table}[htbp]
\scriptsize
\begin{center}
     \begin{tabular}{l l l}
         \toprule
         \textbf{Type}      & \textbf{Name} & \textbf{Description} \\
         \midrule
         \texttt{index\_type} & \texttt{nrows} & Number of rows of the matrix \\
         \texttt{index\_type} & \texttt{ncols} & Number of columns of the matrix \\
         \texttt{offset\_type} & \texttt{nnz} & Structural number of non-zeros \\
         \texttt{offset\_type} array & \texttt{colptr} & Column pointer array (length: \texttt{ncols+1}) \\
         \texttt{index\_type} array & \texttt{rowindx} & Row index array (length: \texttt{nnz}) \\
         \texttt{scalar\_type} array & \texttt{values} & Structural values (length: \texttt{nnz}) \\
         \texttt{base\_type} & \texttt{index\_base} & Base indexing \\
         \bottomrule
     \end{tabular}
\end{center}
     \caption{CSC input parameters}
    \label{tab:csc_param}
\end{table}


\begin{table}[htbp]
\scriptsize
\begin{center}
     \begin{tabular}{l l l}
         \toprule
         \textbf{Type}      & \textbf{Name} & \textbf{Description} \\
         \midrule
         \texttt{index\_type} & \texttt{nrows} & Number of rows of the matrix \\
         \texttt{index\_type} & \texttt{ncols} & Number of columns of the matrix \\
         \texttt{offset\_type} & \texttt{nnz} & Structural number of non-zeros \\
         \texttt{index\_type} array & \texttt{rowindx} & Row index array (length: \texttt{nnz}) \\
         \texttt{index\_type} array & \texttt{colindx} & Column index array (length: \texttt{nnz}) \\
         \texttt{scalar\_type} array & \texttt{values} & Structural values (length: \texttt{nnz}) \\
         \texttt{base\_type} & \texttt{index\_base} & Base indexing\\
         \bottomrule
     \end{tabular}
\end{center}
     \caption{COO input parameters}
    \label{tab:coo_param}
\end{table}

Many sparse formats can be generalized to a block version, leading to more suitable data structures when dealing with dense block structures typical of matrices found in many simulation codes and obtained with the finite element method (FEM) or the discontinuous Galerkin method (DG). These data structures reduce the memory footprint and propose suitable storage for using dense linear algebra.  
For instance, the BSR (or BCSR, Block Compressed Sparse Row) format extends the CSR format by using square block-dense matrices rather than scalars for the values. Information describing the characteristics of its blocks, such as the dimension \texttt{bsize} and the internal storage \texttt{bstorage} (row-oriented or column-oriented layout), must be included. The array \texttt{values} stores non-zero blocks contiguously.

While we acknowledge the wide-spread use of block formats, we exclude them in this first version of an API design proposal. This is motivated by the idea of keeping this document within reasonable length and complexity.

We anticipate using \texttt{std::mdspan}, introduced in \cppstd{23}, to support dense matrices in the same style as the recently standardized \texttt{std::linalg} C++ BLAS interface~\cite{cpplinalg:p1673r13}.  Users can create an \texttt{mdspan} view with a data handle (a pointer is the most common example, although different data handles may appear as defined by an accessor policy) along with matrix extents, representing the matrix's dimensions.  Once an \texttt{mdspan} view has been created, the user can pass in a single argument to represent a dense matrix, instead of passing in multiple arguments for the pointer, dimensions, and stride.  \texttt{std::mdspan} supports a large degree of compile-time customization, including compile-time constant extents, customization of index type view extents, as well as custom layout and accessor policies.  Row- and column- major matrices are supported by standard-defined layout policies (\texttt{std::layout\_right} and \texttt{std::layout\_left}).  We anticipate vendors restricting users to only a subset of the possible \texttt{mdspan}s, such as \texttt{mdspan}s with standard-defined layout policies, default accessors, and an order of one or two.  These \texttt{mdspan}s could be directly passed into most vendor-provided implementations that expect a traditional dense matrix represented by a pointer, dimensions, stride, and row or column major layout.  Unsupported \texttt{mdspan}s would trigger a clear compile error message in a high-quality implementation.

\begin{table}[htbp]
\scriptsize
     \begin{tabular}{l l L{7cm}}
         \toprule
         \textbf{Type}    & \textbf{Name} & \textbf{Description} \\
         \midrule
         \texttt{}
         - & \texttt{T} & Type of values stored in \texttt{mdspan}, equivalent to \texttt{scalar\_type}. \\
         - & \texttt{Extents} & Type specifying the number of dimensions, their types, and whether they are known at compile time or runtime.\\
         - & LayoutPolicy & A type defining how matrix elements are laid out. (e.g. \texttt{std::layout\_left}, \texttt{std::layout\_right})\\
         - & AccessorPolicy & A type defining how matrix elements should be accessed.\\
         \\
         \texttt{data\_handle\_type} & p & Handle to dense matrix values. \\
         \texttt{std::integral} type & \texttt{exts...} & The matrix's extents, or dimensions.\\
         \bottomrule
     \end{tabular}
     \caption{Key template parameters and constructor arguments for building an \texttt{std::mdspan}.  Template parameters, which provide types that customize the \texttt{mdspan}, are shown with ``type'' field empty, while runtime types have their type listed.}
    \label{tab:dense_param}
\end{table}

\section{API Design}
\label{sec:api-design}

The sparse BLAS interface design has to consider a wide range of aspects, ranging from opaqueness to memory allocation for sparse outputs and data ownership to multi-stage APIs and object- and function- specific execution policies and hints for optimization. 
In the following sections, we address these intertwined aspects individually. 

\subsection{Ownership and Opaqueness}

There exist different strategies for handling the ownership of data in sparse linear algebra operations. 
\begin{itemize}
\item A fully owning, less transparent API based on opaque objects.
\item A non-owning, fully transparent API based on the \cplusplus views concept.
\end{itemize}

There are multiple reasons for having these two types of API. 1) transparent, non-owning matrix objects can prevent some vendor optimizations (e.g., using a better-suited matrix format), but on the other hand, owning matrix objects come with an initial usage cost and require copies; 2) non-owning types require multiple phases for their instantiation and allocation which must be explicitly exposed to the user, which can be especially complex and require several steps for more complex operations (e.g., Sparse-Matrix times Sparse-Matrix product -- SpGEMM); 3) with transparent, non-owning objects, the internal data can always be inspected and accessed at little cost, whereas accessors are required for opaque objects and can come with an extra cost.

Because each of these aspects can be important in different situations and to different users, we believe that the two options are both needed, and complementary, and providing these two types of API will satisfy most needs.

In this first version of an interface design, 
we focus on the second option, the non-owning API for functionality. However, as we will detail next, we extend this approach to allow for a moderate level of optimization. 
In this non-owning API, the user owns the data and light-weight views are used to expose the functionality to the library operations. In particular, the light-weight views expose the raw pointers and size information of, for example, a CSR matrix, dense matrix, or a COO matrix.

As discussed, this approach limits the optimization potential for developers. 
In particular, the light-weight views do not allow to 
store additional information that can be exploited
by algorithms operating on the sparse matrices, like symmetry, upper or lower triangular structure, matrix transpose, etc. 
Without breaking the concept of light-weight views, 
we enable optimization and storing additional information by wrapping the view into a \texttt{matrix\_handle} that contains both the view to the user-owned raw data and a structure that can contain additional data for optimization. Important is that all the optimization data is owned by the library, not by the user, and the data is stored in library-allocated memory and is generally opaque to the user.

Some users want to keep control of all available memory, in particular control how data structures are allocated. For this reason, we propose a design where the user can pass an \texttt{allocator} to the library that is then used by the library to allocate additional data in the \texttt{matrix\_handle}. If no allocator is provided, the library is permitted to use whichever allocator they deem to be appropriate.

In \Cref{fig:csrmatrixhandle} we give an example how a CSR matrix structure is passed as a light-weight view to a library and a \texttt{matrix\_handle} is wrapped around it to allow for storing additional information.

\begin{listing}[H]
\begin{minted}{c++}
using namespace sparseblas;
csr_view<float> A(values, rowptr, colind, shape, nnz);

// matrix_handle is opaque and may contain 
// vendor optimization details
auto A_handle = matrix_handle(A, allocator);
\end{minted}
\label{fig:csrmatrixhandle}
\caption{Wrapping a CSR matrix structure composing of \texttt{values, rowptr, colind, shape, nnz} into a \texttt{matrix\_handle} with user-specified \texttt{allocator}.}
\end{listing}

We note that the API is expected to accept both plain views or \texttt{matrix\_handle}'s. If plain views are passed in, the library has no optimization structure at hand and can not store any additional matrix information that persists beyond the operation.

\subsection{Horizontal and Vertical Interoperability}

The interface and functionality standardization we propose must allow for horizontal and vertical interoperability. 
\textit{Vertical interoperability} includes both upward and downward interoperability. Upward interoperability refers to usability and productivity inside high-level algorithms like iterative solvers, preconditioners, and sparse neural networks. Only interfaces and functionality decisions allowing for high productivity for high-level algorithms and applications justify the design of a sparse BLAS standard. We assess upward interoperability by using the sparse BLAS functionality inside mini apps. 
Downward interoperability refers to the possibility of wrapping already existing sparse BLAS functionality developed by hardware vendors into the sparse BLAS interface. The intention for the downward interoperability is not to create yet another layer of interfaces, but to ensure that design decisions taken by the vendors in the kernel implementations do not hinder the adoption of a new interface in future kernel 
development.
\textit{Horizontal interoperability} refers to compatibility with dense
BLAS functionality and the possibility to use the functionality in other
programming languages. The latter takes on a new significance in the
context of Sparse BLAS, as we aim to focus on building a high-level
\cplusplus API, as opposed to the Fortran-based interface used in the
dense BLAS to allow for ABI compatibility. Instead, our goal is to ensure that basic functionality is accessible from a wide range of languages, including Fortran, C, Julia, Python, and others. The strategy we envision is to create C-bindings for the most central functionality.

Most vendors implement linear algebra functionality using the
\cplusplus programming language. Likewise, most already implemented
sparse BLAS functionality, e.g. in NVIDIA's, Arm's, Intel's, and AMD's
vendor libraries, are often designed in \cplusplus. Also, for dense
BLAS, the \cplusplus standard officially adopted the
\texttt{std::linalg} API~\cite{cpplinalg:p1673r13}. We adopt this
strategy and build upon an interface in \cplusplus. At the same time,
we acknowledge that many users and application developers depend on Fortran, C, or Python interfaces as their application codes are using the respective language. Designing an interface to sparse BLAS functionality in C or Fortran without the use of static and dynamic polymorphism is almost impossible due to the explosion in variants, which would result in a prohibitively long parameter list and a plethora of function variants. In response, we will select the most common use cases and provide bindings in C, Fortran, and Python to a subset of the sparse BLAS functionality.

\subsection{Review of existing API designs SpMV }
\label{sec:spmvReview}
While the community has not followed a synchronized approach towards an API for sparse linear algebra functionality, several groups from academia and industry have developed and deployed sparse linear algebra operations along with a user-facing interface. While the interfaces differ, they are all designed to accommodate functionality- and hardware-specific needs. Most of the currently supported APIs follow the transparent, non-data-owning design. 
Existing APIs are different but share certain design patterns, and we consider them as blueprints guiding the design of the new API standard for sparse linear algebra functionality.
Thus, in the subsequent, we provide a comparison of existing C and
\cplusplus library APIs for sparse matrix-vector (SpMV) operations.
Some APIs are synchronous and blocking, while others extend to asynchronous non-blocking execution models.

\newpage
\noindent\textbf{SpMV API comparisons between existing libraries:}\\ IEEE754 Double Precision Sparse Matrix-Vector Multiplication (SpMV) \begin{equation*}y = \alpha\cdot\text{op}\left(A\right) x + \beta\cdot y\end{equation*}
\begin{enumerate}[label=(\arabic*)]
\item Arm Performance Libraries C API (has an optional hint/optimize stage)
\begin{lstlisting}
// comm
/* comm */
info = armpl_spmv_exec_d(ARMPL_SPARSE_OPERATION_NOTRANS, alpha, armpl_mat, x, beta, y);
\end{lstlisting}
\item oneMKL C API (has an optional hint/optimize stage)
\begin{lstlisting}
status = mkl_sparse_d_mv(SPARSE_OPERATION_NON_TRANSPOSE,
            alpha, csrA, descrA, x, beta, y);
\end{lstlisting}
\item AOCL-sparse C API (has an optional hint/optimize stage)
\begin{lstlisting}
status = aoclsparse_dmv(aoclsparse_operation_none,
            p_alpha, A, descr_a, y, p_beta, x);
\end{lstlisting}
\item Graph BLAS Suite Sparse C API (implements $y\left<\text{mask}\right> = \text{accum}\left(y, A\cdot x\right)$ in some semiring)
\begin{lstlisting}
// y = alpha*y
GrB_apply(y, GrB_NULL, GrB_NULL, GrB_TIMES_FP64, alpha, y, GrB_NULL);
y += beta*A*x
GrB_mxm (y, GrB_NULL, GrB_PLUS_FP64. GrB_PLUS_TIMES_SEMIRING_FP64, A, x, GrB_NULL);
\end{lstlisting}
\item ALP/GraphBLAS \cplusplus (minimum number of containers, optimised by nonblocking execution)
\begin{lstlisting}
grb::Semiring< grb::operators::add<double>, grb::operators::mul<double>,
    grb::identities::zero, grb::identities::one > plusTimes;
grb::Monoid< grb::operators::mul< double >, grb::identities::one > mul;
status = grb::foldr( beta / alpha, y, mul );
status = status ? status : grb::mxv< op >( y, A, x, plusTimes );
status = status ? status : grb::foldr( alpha, y, mul );
\end{lstlisting}
\item cuSPARSE CUDA C, generic API (has multiple required stages)
\begin{lstlisting}
status = cusparseSpMV(handle, CUSPARSE_OPERATION_NON_TRANSPOSE,
            p_alpha, matA, vecX, p_beta, vecY,
            CUDA_R_64F, CUSPARSE_SPMV_ALG_DEFAULT,
            externalBuffer);
\end{lstlisting}
\item rocSPARSE ROCm C, generic API (has multiple required stages)
\begin{lstlisting}
status = rocsparse_spmv(handle, rocsparse_operation_none,
            p_alpha, matA, vecX, p_beta, vecY,
            rocsparse_datatype_f64_r,
            rocsparse_spmv_alg_csr_adaptive,
            rocsparse_spmv_stage_compute,
            p_buffer_size, temp_buffer);
\end{lstlisting}
\item hipSPARSE ROCm/CUDA C API
\begin{lstlisting}
status = hipsparseDcsrmv(handle, HIPSPARSE_OPERATION_NON_TRANSPOSE,
            m, n, nnz, p_alpha, descrA,
            valA, rowptrA, colindA, x, p_beta, y);
\end{lstlisting}
\item oneMKL SYCL \cplusplus API with an out-of-order queue (has an optional optimize stage)
\begin{lstlisting}
ev_gemv = oneapi::mkl::sparse::gemv(
            queue, oneapi::mkl::transpose::nontrans,
            alpha, csrA, x, beta, y, {ev_opt});
\end{lstlisting}
\item oneMKL SYCL \cplusplus API with an in-order queue (has an optional optimize stage)
\begin{lstlisting}
oneapi::mkl::sparse::gemv(
            queue, oneapi::mkl::transpose::nontrans,
            alpha, csrA, x, beta, y);
\end{lstlisting}
\item Kokkos-kernel \cplusplus API (can set algorithm in \textit{controls} argument)
\begin{lstlisting}
KokkosSparse::spmv(controls, KokkosSparse::NoTranspose,
            alpha, A, x, beta, y);
\end{lstlisting}
\item Ginkgo \cplusplus API
\begin{lstlisting}
// cuda/rocm stream or sycl::queue is attached to A previously
// (through an "Executor" object)
// matrix::Csr *A;              // sparse matrix
// matrix::Dense *y, *x;        // vectors tall and skinny matrices
// matrix::Dense *alpha, *beta; // scalars of dimension 1x1
A->apply(alpha, x, beta, y);
\end{lstlisting}

\end{enumerate}

Note that the \textit{handle} in rocSPARSE, cuSPARSE and hipSPARSE APIs is an object that should be created with a call to \verb|XYZsparseCreate(p_handle)| before any other API is invoked at which time it globally initializes their Sparse BLAS library and creates the \textit{handle} on their device \textit{context}. It also houses the appropriate \verb|cudaStream_t|, \verb|rocStream_t| or \verb|hipStream_t| \textit{stream} to be used in subsequent calls.  The CPU APIs have no such wrapper, and the SYCL APIs take a \textit{sycl::queue}.

\subsection{API Design Considerations}
From a high-level point of view, there exist two classes of sparse BLAS operations:
\begin{enumerate}
    \item Operations where the sparsity pattern of the output object is known a priori, e.g. the multiplication of a sparse matrix with a vector or a dense matrix, or the sampled multiplication of two dense matrices, or the scaling of a matrix. These operations can be realized with a single kernel operating on pre-allocated output data.
    \item Operations where the sparsity pattern of the output object is not known a priori, e.g. the multiplication of two sparse matrices, a filtering operation removing values below a pre-defined threshold, or the addition of two sparse matrices (with -- in general -- different sparsity patterns).
    For these operations, it is impossible to execute them using a single kernel as the memory requirement for the output is not known a priori, and the memory allocation can only happen after the number of nonzero elements has been computed. For this reason, operations of this class are typically implemented using the pattern \textit{compute}, \textit{allocate}, \textit{fill}~\cite{10.1145/275323.275327}.
\end{enumerate} 

For both single-stage and multi-stage operations that produce sparse output, the sparsity structure of the output is based on the symbolic operation, not on the numeric operations. In other words, a numeric calculation resulting in a numeric zero does not result in omitting this entry from the output sparsity pattern.

While there exist strong arguments for unifying the interface for the different functionality to always require the same number of stages (conceptually, a matrix matrix multiplication is the same operation, agnostic to the non-zero structure and the storage format of the matrices), we decided to sacrifice consistency in favor of reduced complexity. In particular, we use a one-stage API for functionality that produces output of known sparsity pattern, and a multi-stage API for functionality that produces output of unknown sparsity pattern. 

Functionality operating on sparse objects differ also in another aspect from functionality operating on dense objects: depending on the distribution of the nonzero entries, different parameter choices may be optimal for fast execution -- up to the level where different algorithms may be chosen depending on the nonzero distribution. To account for this, we consider an additional \textit{inspect phase} that is optional to all functionality (single-stage or multi-stage) and allows for additional optimization with regard to the selection of the algorithm and parameters for processing the operation and for storing some additional information in the matrix handle. The optimization may in particular pay off in case of repeated function invocation, e.g., the repeated invocation of a sparse matrix vector product inside an iterative solver loop. If the user feeds in plain views into the operation, the inspect phase can invoke operation-specific optimization but can not access additional information in the optimization structure of a matrix handle.
If the optional inspect phase is not invoked prior to the operation, a default algorithm and parameter selection is used. 
This concept of an optional inspect phase is in line with the current design of vendor interfaces, see Section \ref{sec:spmvReview}.

In the next sections, we will discuss the single-stage API for functionality producing output with a-priori known sparsity pattern and the multi-stage API for functionality producing output with a-priori unknown sparsity pattern.
Note, that all operations take \texttt{policy} and \texttt{state} objects as inputs. The \texttt{policy} object is opaque to the user. It may contain any vendor-specific execution policies, similar to a \texttt{stream} or \texttt{queue}. The \texttt{state} object is operation-specific and can be used to store any operation-specific data, e.g. a dependency graph in a triangular solve or data in an operation-specific library-internal representation. For the multi-stage functionality, it may also contain information about which stages have already been completed to avoid re-calculation.

\subsection{Single-Stage API for Operations with known Output}
\label{sec:single-stage-api}

Operations with known output can be realized using a single routine. Functionality listed in Table~\ref{tab:scope} falling into this category includes: scaling, transpose and conjugate transpose, norms, sparse matrix vector multiplication, sparse matrix dense matrix multiplication, triangular solve, general and sampled dense matrix multiplication. As previously mentioned, we allow for an optional \textit{inspect} phase for inspecting the sparse objects and passing information to the routines executing the operation. We here present the interface for some of the functionality, noting that the functionality in general accepts both, plain views on the objects and matrix-handles, with the latter allowing for additional optimization.


\begin{listing}[H]
\begin{minted}{c++}
using namespace sparseblas;
csr_view<float> A(values, rowptr, colind, shape, nnz);
// auto A = matrix_handle(csr_view<float>(...), allocator);

// scale() overwrites the values of A 
scale_state_t state(allocator);
scale(policy, state, 2.3, A);
\end{minted}
\caption{Scaling, $A := \alpha A$}
\end{listing}


\begin{listing}[H]
\begin{minted}{c++}
using namespace sparseblas;
csr_view<float> A(values, rowptr, colind, shape, nnz);

// matrix_handle is opaque and may contain 
// vendor optimization details
auto A_handle = matrix_handle(A, allocator);

// A is const; function returns Inf norm
matrix_inf_norm_state_t state(allocator);
auto inf_nrm = matrix_inf_norm(policy, state, A);
\end{minted}
\caption{Inf Matrix Norm, $\alpha = \|A\|_\text{inf}$.}
\end{listing}


\begin{listing}[H]
\begin{minted}{c++}
using namespace sparseblas;
csr_view<float> A(values, rowptr, colind, shape, nnz);
// auto A = matrix_handle(csr_view<float>(...), allocator);

matrix_frob_norm_state_t state(allocator);
// A is const; function returns Frobenius norm
auto frob_nrm = matrix_frob_norm(policy, state, A);
// or - without passing a policy, resulting in a default:
auto frob_nrm = matrix_frob_norm(state, A);
\end{minted}
\caption{Frobenius Matrix Norm, $\alpha = \|A\|_F$}
\end{listing}

\begin{listing}[H]
\begin{minted}{c++}
using namespace spblas;
csr_view<float> A(values, rowptr, colind, shape, nnz);

auto x = std::mdspan(raw_x.data(), 100);
auto y = std::mdspan(raw_y.data(), 100);

// matrix_handle is opaque and may contain 
// vendor optimization details
auto A_handle = matrix_handle(A, allocator);

// create a state for the optimization of the routine
multiply_state_t state(allocator);

// optional inspect phase can add information 
// to the matrix handle
multiply_inspect(policy, state, A_handle, x, y);

// actual multiplication y = A * x
multiply(policy, state, A_handle, x, y); 

// alternatively, the multiplication can use the plain views
// limiting the optimization opportunities
multiply(policy, state, A, x, y); 
\end{minted}
\caption{Sparse matrix vector product $y = A \cdot x$.}
\end{listing}

\begin{listing}[H]
\begin{minted}{c++}
using namespace spblas;
csr_view<float> A(values, rowptr, colind, shape, nnz);
float alpha = ..., beta = ...;

auto X = std::mdspan(raw_x.data(), 100, 100);
auto Y = std::mdspan(raw_y.data(), 100, 100);

// matrix_handle is opaque and may contain
// vendor optimization details
auto A_handle = matrix_handle(A, allocator);

// create a state for the optimization of the routine
multiply_state_t state(allocator);

// optional inspect phase can add information
// to the matrix handle
multiply_inspect(policy, state, scaled(alpha, A_handle), X, scaled(beta, Y), Y);

// actual multiplication Y = alpha * A * X + beta * Y
multiply(policy, state, scaled(alpha, A_handle), X, scaled(beta, Y), Y); 

// alternatively, the multiplication can use the plain views
// limiting the optimization opportunities
multiply(policy, state, scaled(alpha, A), X, scaled(beta, Y), Y); 
\end{minted}
\caption{Sparse matrix dense matrix product, $Y = \alpha \cdot A \cdot X + \beta \cdot Y$, with $X$ and $Y$ being dense matrices.}
\end{listing}

\begin{listing}[H]
\begin{minted}{c++}
using namespace spblas;
// Matrix T data is already in lower or upper triangular form
// physically in the matrix arrays for the triangular
// solve
csr_view<float> T(values, rowptr, colind, shape, nnz);

auto x = std::mdspan(raw_x.data(), 100);
auto b = std::mdspan(raw_y.data(), 100);

// matrix_handle is opaque and may contain
// vendor optimization details
auto T_handle = matrix_handle(T, allocator);

// create a state for the optimization of the routine
triangular_solve_state_t state(allocator);

// optional inspect phase can add information
// to the matrix handle
triangular_solve_inspect(policy, state, T_handle, b, x);

// actual triangular solver x = T \ b
triangular_solve(policy, state, T_handle, b, x); 

// alternatively, the triangular solver can use the plain views
// limiting the optimization opportunities
triangular_solve(policy, state, T, b, x); 
\end{minted}
\caption{Triangular solve: solve $T\cdot x = b$ for x.}
\end{listing}

\begin{listing}[H]
\begin{minted}{c++}
using namespace sparseblas;
auto X = std::mdspan(raw_x.data(), m, k);
auto Y = std::mdspan(raw_b.data(), k, n);
// sample mask is the pattern of CSR matrix output
csr_view<float> C(values, rowptr, colind, shape, nnz);

sampled_multiply_state_t state(allocator);
sampled_multiply_inspect(policy, state, X, Y, C);

// C = C.*( X * Y ): vendor optimizations in state for reuse
sampled_multiply(policy, state, X, Y, C); 

\end{minted}
\caption{Sampled dense dense matrix multiplication (SDDMM), $C\langle \text{mask} \rangle = X\cdot Y$, where the $C$ sparsity pattern encodes the mask}
\end{listing}

\subsection{Multi-Stage API for Operations with unknown Output}
\label{sec:multi-stage-api}

As mentioned above, a fundamental difference to dense BLAS functionality is that for basic operations on sparse objects, the sparsity pattern (and therewith the sparse data structure) is not always known beforehand. An example is the multiplication of two sparse matrices $A, B$, resulting in another sparse matrix $C = A\cdot B$ with a pattern that typically is unknown prior to the computation. In consequence, functionality producing sparse output with the sparsity pattern unknown prior to the computation typically split into three phases (or four if including the optional inspect phase):
\begin{enumerate}
    \item \textit{inspect phase} -- (optional) prepare any potential optimizations for subsequent phases, or may do nothing
    \item \textit{compute phase} -- computing the size of the sparse output structure, which typically requires significant work
    \item \textit{allocation phase} -- allocating the memory for the sparse output data structure and placing it in the output matrix object to be filled
    \item \textit{fill phase} -- complete execution of the operation and filling the output structure with the result.
\end{enumerate}

An important design question for functionality producing sparse output is whether the memory allocation is done by the user or the library. While there exist strong arguments on both sides, and passing a memory allocator to the library poses an elegant way to resolve memory location and ownership, we initially propose an interface that requires the library user to allocate the memory for the sparse output data after providing the memory requirement information. While we, in the following, focus on the multiplication of sparse matrices (typically called SpGEMM in literature), the considerations and design aspects hold also for other basic functionalities that are producing output with sparsity pattern unknown prior to the computation. In Table~\ref{tab:scope} we list several: sparse matrix -- sparse matrix multiplication, matrix sparse format conversion, filter operations removing elements below a pre-defined threshold, and the addition of sparse matrices with generally differing sparsity patterns.
 
Acknowledging the previously motivated \textit{inspect} phase, the multiplication of two sparse matrices 

\begin{equation*}
C = \alpha \cdot \text{op}(A) \cdot \text{op}(B) + \beta \cdot D
\end{equation*}

can be realized using the following sequence of steps (using APIs that will be defined subsequently):

\begin{minted}{c++}
using namespace spblas;
// csr_view<float> A,B,D filled elsewhwere
// csr_view<float> C(c_nrows, c_ncols);

sparse_multiply_state_t state(allocator);
sparse_multiply_inspect(policy, state, A, B, C, D); // optional
sparse_multiply_compute(policy, state, A, B, C, D);
index_t nnz = state.get_result_nnz();
// the user allocates the arrays for C
sparse_multiply_fill(policy, state, A, B, C, D);
// C structure and values are now ready to be used
\end{minted}
We note that we change from the function name \texttt{multiply} (which we used for the multiplication of sparse with dense matrices or vectors) to \texttt{sparse\_multiply} to indicate that this operation generates sparse output and that a fundamentally different multi-stage API is needed. In consequence, it is not possible to just call \texttt{multiply(A,B,C)} on sparse objects \texttt{A,B,C}.

We also note that this interface gives the software developers a significant level of flexibility and room for optimization. 

A standard way to is to place the structural analysis in the \texttt{multiply\_compute} stage, then allocate the output data structure and fill it in the \texttt{multiply\_fill} stage. Internal buffers needed in the structural analysis can be allocated and de-allocated by the library in the \texttt{multiply\_compute} routine. 

The only step that has to critically happen in the \texttt{multiply\_compute} routine is the computation of the number of nonzero entries in the output to be able to allocate the output data structure. The \texttt{multiply\_fill} routine is required to fill the user provided arrays with the output data so that it is usable after \texttt{multiply\_fill} stage. These are the only requirements on the \texttt{multiply\_compute} and \texttt{multiply\_fill} stages, and thus the developers are left with a lot of freedom to realize the functionality. For example, developers can decide to already fill the CSR \texttt{rowptr} in the \texttt{multiply\_compute} routine. Or pack even more into this first stage: An approach allowing for more library-internal optimization may be to pack both the structural analysis and the multiplication into the \texttt{multiply\_compute} routine using an internal output data structure that is part of the \texttt{state} object. The \texttt{multiply\_fill} routine then only extracts the result into the output data structure allocated by the user. Any library-internal output data structures live in the state object and would be deleted when state is destroyed. Such an implementation does not break the interface design.

A critical question is how the library allocates these internal buffers, via a library-preferred allocator or an allocator passed to the library from the user. This boils down to whether the user has buffer management on the application level, i.e., manages all the device memory in the application. At the same time, requiring the user to always provide an allocator to the library can become cumbersome. In the current design proposal, we allow the user to pass in a memory allocator (for the memory space where the computations are executed, i.e. the ``compute device'') that is then used by the library for the allocation of buffers. If no allocator is passed, the library is free to use its own allocator.

Some scenarios additionally benefit from splitting the multiplication of sparse matrices and other routines operating on sparse data structures into a symbolic and a numeric stage. This, for example, can allow for additional optimization in the repeated multiplication of matrices changing only the numeric values but preserving the sparsity pattern. Hence, we propose a variant of the above interface splitting the routines
\texttt{multiply\_compute} and
\texttt{multiply\_fill} into two:

\begin{minted}{c++}
using namespace spblas;
// csr_view<float> A,B,D filled elsewhwere
// csr_view<float> C(c_nrows, c_ncols);

sparse_multiply_state_t state(allocator);
multiply_inspect(policy, state, A, B, C, D); // optional
multiply_symbolic_compute(policy, state, A, B, C, D);
index_t nnz = state.get_result_nnz();
// allocate C arrays and put in C (possibly both structure
// and values or just structure arrays at this point)
multiply_symbolic_fill(policy, state, A, B, C, D);
// C structure is now able to be used

// allocate, if necessary, values arrays and place them in C
multiply_numeric_compute(policy, state, A, B, C, D);
multiply_numeric_fill(policy, state, A, B, C, D);
// C structure and values are now able to be used
\end{minted}

Because we do not specify at which phase the output matrix is prepared internally, the \textit{compute phase} (e.g. \texttt{multiply\_numeric\_compute}) and the \textit{fill phase} (e.g. \texttt{sparse\_multiply\_numeric\_fill}) should be regarded as a pair and for repeated execution, both phases should be called.  It is up to implementations to detect on repeated calls which internal data in the \texttt{multiply\_state\_t} object is reusable and what must be recomputed.

We want to provide some examples for the use of this API. 
\begin{enumerate}
    \item Single use:
\begin{minted}{c++}

sparse_multiply_state_t state(allocator);
sparse_multiply_inspect(policy, state, A, B, C, D); // optional
sparse_multiply_compute(policy, state, A, B, C, D);
index_t nnz = state.get_result_nnz();
// allocate C arrays 
sparse_multiply_fill(policy, state, A, B, C, D);
// C structure is now able to be used
\end{minted}
\item 
Repeated invocation of the functionality with numerically differing matrices:
\begin{minted}{c++}
sparse_multiply_state_t state(allocator);
sparse_multiply_inspect(policy, state, A, B, C, D); // optional
for (i=0 i<n; i++) {
  // may do nothing if nnz is already computed
  // and information is stored in the state
  sparse_multiply_compute(policy, state, A, B, C, D);
  index_t nnz = state.get_result_nnz();
  if (i == 0 }{
    // allocate C arrays 
  }
  sparse_multiply_fill(policy, state, A, B, C, D);
  // C structure is now able to be used
}
\end{minted}
\item 
Using the split into numeric and symbolic phase for repeated invocation of the functionality with numerically differing matrices:
\begin{minted}{c++}

sparse_multiply_state_t state(allocator);
sparse_multiply_inspect(policy, state, A, B, C, D); // optional
sparse_multiply_symbolic_compute(policy, state, A, B, C, D);
index_t nnz = state.get_result_nnz();
// allocate C arrays 
sparse_multiply_symbolic_fill(policy, state, A, B, C, D);
for (i=0 i<n; i++) {
  // only numeric computations inside the loop
  multiply_numeric_compute(policy, state, A, B, C, D);
  multiply_numeric_fill(policy, state, A, B, C, D);
  // C structure is now able to be used
}
\end{minted}
\end{enumerate}

Up to now, we provided motivation and details for the design of the API for the sparse matrix - sparse matrix multiply operation, but the same two-stage interface applies also to other functionality with nonzero structure unknown a-priori.

\begin{listing}[H]
\begin{minted}{c++}
using namespace sparseblas;
auto A = std::mdspan(raw_x.data(), m, k);
csr_view<float> B(m, n);

convert_state_t state(allocator);
convert_inspect(policy, state, A, B);
convert_compute(policy, state, A, B);
index_t nnz = state.get_result_nnz();
// allocate the memory for B and put them in B
convert_fill(policy, state, A, B); 
\end{minted}
\caption{Sparse Matrix Format Conversion, $B = \text{sparse}(A)$. Note that this will not remove explicit zeros except for conversion from dense. The above convert\_* functions might accept the predicate function additionally to remove some entries during conversion, which can give some performance benefit without creating a matrix two times.}
\end{listing}


\begin{listing}[H]
\begin{minted}{c++}
using namespace sparseblas;
csr_view<float> A(values, rowptr, colind, shape, nnz);
// auto A = matrix_handle(csr_view<float>(...), allocator);
// ... likewise for B
csr_view<float> C(m, n);
// auto C = matrix_handle(csr_view<float>(...), allocator);

multiply_elementwise_state_t state(allocator);
multiply_elementwise_inspect(policy, state, A, B, C); // optional
multiply_elementwise_compute(policy, state, A, B, C);
index_t nnz = state.get_result_nnz();
// allocate C arrays and put in C
multiply_elementwise_fill(policy, state, A, B, C);
// C structure and values are now able to be used
\end{minted}
\caption{Element-wise Multiplication, $C = A~.*B$.}
\end{listing}


\begin{listing}[H]
\begin{minted}{c++}
using namespace sparseblas;
csr_view<float> A(values, rowptr, colind, shape, nnz);
// auto A = matrix_handle(csr_view<float>(...), allocator);
// ... likewise for B
csr_view<float> C(m, n);
// auto C = matrix_handle(csr_view<float>(...), allocator);

add_state_t state(allocator);
add_inspect(policy, state, A, B, C); // optional
add_compute(policy, state, A, B, C);
index_t nnz = state.get_result_nnz();
// allocate C arrays and put in C
add_fill(policy, state, A, B, C);
// C structure and values are now able to be used
\end{minted}
\caption{Sparse Matrix -- Sparse Matrix Addition, $C = A + B$.}
\end{listing}

\begin{listing}[H]
\begin{minted}{c++}
using namespace sparseblas;
csr_view<float> A_view(values, rowptr, colind, shape, nnz);
csr_view<float> B(values, rowptr, colind, shape, nnz);

auto pred = [](auto i, auto j, auto v) {
              return v > 0;
            };

auto pred = [](auto i, auto j, auto v) {
              return (i < 10 ) && (j < 10);
            };

filter_state_t state(allocator);
           
// matrix_handle is opaque and may contain vendor optimization details
matrix_handle A(A_view, allocator);
filter_compute(policy, state, A, B, pred);
index_t nnz = state.get_result_nnz();
// the user can allocate the arrays for the output structure
// finally, the output structure can be filled
filter_fill(policy, state, A, B, pred); 
\end{minted}
\caption{Predicate Selection, $B = A.*(A>0)$.}
\end{listing}


\newpage

\newpage

\noindent


\section{Execution Model}
\label{sec:async_api}

The world of high-performance computing has recently transformed from a CPU-centric model to a heterogeneous model in which computations must be performed on CPUs, GPUs, and other accelerators. To support these heterogeneous environments, computing languages and libraries have evolved from supporting only \textit{synchronous execution models}, where a kernel or function is submitted, enqueued, and executed immediately to a \textit{asynchronous execution models}, where kernel submission is executed without waiting for completion. This separation between kernel submission and kernel execution has proved beneficial for enabling performance on heterogeneous systems by both avoiding unnecessary synchronization as well as allowing data movement, kernel launch, and other overheads associated with heterogeneous execution to be hidden.

Standard \cplusplus is transitioning toward a
standard model for asynchronous execution with \texttt{std::execution} (senders and receivers)~\cite{p2300r7,p3300r0}, serves as the glue for different libraries with asynchronous operations to interoperate with one another.  Given the increased performance and ease-of-use associated with using standard tools to handle asynchrony instead of disparate vendor APIs, we recognize the importance of using standard, interoperable tools for asynchrony. In the future, we anticipate adopting the \texttt{std::execution} model, with a collection of asynchronous methods that extend the standard Sparse BLAS API using \texttt{std::execution} concepts.  However, given the early stage of \texttt{std::execution} implementations, this full specification will likely have to wait until the ecosystem is more mature.

With the current state of vendor support for asynchronous C++ programming, we anticipate four guidelines at this stage for asynchronous support.
First, any Sparse BLAS API that supports asynchronous execution should be included in the \verb|async| namespace ~\cite{p3300r0}, to clearly denote expected behavior for users. For example, \verb|spblas::async::multiply()|.
Second, all asynchronous and synchronous APIs that inspect or execute shall take an \textit{execution policy} as an argument and respect it with regards to the parallel execution. In addition to the standard execution policies defined by \cplusplus, vendors may define their own execution policies, which have implementation-defined behavior. Vendors should also provide APIs that do not require an execution policy.  These execution policy--free APIs may perform execution in parallel so long as they do not violate the semantic requirements of their respective algorithms.
Third, any implementation-defined objects that control asynchronous behavior (e.g. CUDA/ROCm \textit{stream}s or SYCL \textit{sycl::queue} and \textit{sycl::event} objects) can be provided to the algorithm through an implementation-defined execution policy.
Fourth, we observe that scaling values in Sparse BLAS operations (e.g. $\alpha$ and $\beta$ in sparse matrix-vector multiplication, $y = \alpha A\cdot x + \beta y$) may be represented as either arguments provided on the host or as device-side scalars that will be dynamically computed by another asynchronous kernel.  A high-quality implementation should support both sources of scaling values.

Some libraries provide a mechanism for the user to select which of one or more internal algorithms they want to be used in a particular sparse BLAS operation. The default behavior is for the library to make such algorithmic choices, perhaps using information gained from analysis of the inputs in the optional ``inspect phase.'' However, in the case that the user would like to select a particular implementation-defined algorithm implementation, this should be supported by vendors via their own implementation-defined execution policies.

\section{Numerical considerations}


This section describes the design space impacting the
numerical properties of the Sparse BLAS, outlines the
choices to be made, explains how these choices may tradeoff performance and ``consistency'' (to be defined below), 
and proposes which options should be required, recommended, 
or not recommended. 
The design space is defined by the following, often intertwined considerations:

\begin{enumerate}
    \item
    Which floating point formats are supported as input/output and 
    which format is used internally for arithmetic operations?
    The growing variety of short (16-bit and 8-bit) formats, 
    block floating point, as well as
    mixed precision (e.g. using a higher or lower precision internally) suggests a design that allows for high flexibility. We describe this design space in more
    detail below, but our initial design just proposes
    standard IEEE 754 32-bit and 64-bit formats, pending user feedback requesting more
    options.
    \item What error bounds should be satisfied in the
    absence of arithmetic exceptions? This is again 
    motivated by the growing variety of available precisions.
    We outline all the possible error bounds below, but again
    our initial design will again recommend just standard error bounds associated with
    IEEE 754 formats.
    \item How should floating-point exceptions (Inf, NaN) be handled?
    Parts of this design consideration are covered by the 
    dense BLAS discussions \cite{Proposed_BLAS_LAPACK_exception_handling}. But other aspects are unique
to the Sparse BLAS because of the impact of sparse data structures. For example, converting an implicit zero in
a sparse matrix to an explicit zero (e.g., as part of a
SIMD optimization) may change how exceptions are 
propagated, i.e., whether an implicit $A(i,j)=0$ is
multiplied by $x(j) = \text{Inf}$ or not when computing
$y(i) = \sum_{k=1}^n A(i,k) \cdot x(k)$. We discuss
different levels of consistency with which exceptions
can be handled. Our initial design does not require any particular way to handle
sparsity, to allow implementations maximum opportunities for performance
optimization.
\item Do we expect bitwise reproducibility of results from run to run?
Floating point addition is not associative, so users
cannot expect bitwise identical results from different
implementations on different platforms, or even from
run to run of the same implementation on the same
parallel platform. Reproducibility has been identified
as a need for various reasons, such as debugging, and is provided
by Intel MKL with CNR (Conditional Numerical
Reproducibility).
This is a higher, and more expensive, 
level of consistency than just exception handling. So again, our initial design
does not require reproducibility.
\item Do we need test suites? We describe how extensively test code
should test a proposed SparseBLAS implementation, given
the design considerations above.
\end{enumerate}

\subsection{Choices of numerical formats}

We begin by listing the wide and growing variety of
numerical formats that are available or will be soon.
This recent growth is motivated by modern AI/ML algorithms
that can use lower precision to get accurate answers much
faster, and with less memory and energy, than using 
traditional 32 and 64-bit floating point formats \cite{IEEE754}. We note that the IEEE 754 Floating Point Standard \cite{IEEE754} also defines 16 and 128-bit formats.

Industry and academia have introduced several low-precision data types in recent years, including TensorFloat32~\cite{ampere}, 4-bit, 8-bit and 16-bit floating point~\cite{micikevicius2022fp8}, Bfloat16~\cite{wang2019bfloat16}, micro-scaling (MX) variants~\cite{rouhani2023microscaling}, which are variations
on classical block floating point \cite{Wilkinson63},
and posits, which have variable precision \cite{Posits,GoodBadUgly}.
Also, a new IEEE committee has recently convened to try to
standardize 8-bit floating point, since several companies
have been building their own separate, incompatible versions \footnote{https://github.com/P3109/Public}.
Vendors are currently prioritizing the acceleration of specific operations, notably general matrix multiplication (GEMM) utilizing low- and mixed-precision cores that are now available on newer hardware, including NVIDIA's Tensor Cores~\cite{ampere}, 
AMD's Matrix Cores~\cite{amd}, IBM's MMA~\cite{starke2021ibm},
and Arm's SME~\cite{sme}, among other specialized architectures~\cite{NorriePYKLLYJP21,cerebras}.
Accelerators equipped with these lower- and mixed-precision cores outperform traditional CPUs/GPUs in AI, ML, and HPC workloads and are more power efficient~\cite{Atoofian23}.
These advantages are in turn motivating the scientific
computing community to use them when possible to accelerate
their higher precision algorithms, by using mixed precision
\cite{higham2022mixed, haidar2020mixed}.

It is worth distinguishing two kinds of numerical formats,
those where all the data needed to interpret a number
is stored in a contiguous set of bits, and those where a
separate scaling factor is needed for a block of numbers.
The former includes the most commonly used floating point formats,
and the latter includes the traditional block floating point, 
micro-scaling variants, and the possible new 8-bit IEEE standard
mentioned above. Having separate scaling factors,
that may apply to different subsets of the input
and output arguments,
will require more complicated data structures, which
requires further discussion to attain portability.

For the Sparse BLAS API, we use template parameters for the input and output formats, 
and anticipate the support for IEEE 754 32- and 64-bit formats, for both real and
complex data. We acknowledge that the format used for the internal arithmetic operations
might be different, i.e. of higher precision to reduce accumulation errors. In particular,
some architecture units already implement a higher accumulation format in hardware. 
In the Sparse BLAS API, we do not specify the internal precision, but anticipate it having at least the accuracy of the input/output format. We anticipate that any deviation from this concept is documented.
We encourage the implementation of functionality variants,
as long as also these are well-documented (including their numerical properties).

\subsection{Error bounds to be satisfied}
\label{Error_Bounds}

With the emergence of various arithmetic formats, and using mixed precision,
it is becoming increasingly important for numerical software to carefully
document what accuracy and related properties users can expect 
(in the absence of numerical exceptions, which are discussed 
in the next subsection).

Based on existing formats and
algorithms, we propose a framework for describing error bounds which
should cover a variety of the precisions described in the last
subsection, including mixed precision, and give several examples.
For simplicity, we consider just the error of computing a
single dot product $z = x^T \cdot y = \sum_{i=1}^n x_i \cdot y_i$.

There are three possible sources of error that an error bound 
needs to take into account: converting the input format(s) of 
$x$ and $y$ to the internal format(s) of $\hat{x}$ and $\hat{y}$ 
in which the arithmetic is done, computing 
the dot product $\sum_{i=1}^n \hat{x}_i \cdot \hat{y}_i$ 
in the internal format, and converting the result $\hat{z}$
to the output format, yielding the final approximation 
$\hat{\hat{z}}$. It is most straightforward to bound
the maximum absolute error that each step can have, and sum them,
yielding the bound
\begin{eqnarray*}
|z - \hat{\hat{z}} | & \leq &  
| \sum_{i=1}^n x_i \cdot y_i - \sum_{i=1}^n \hat{x}_i \cdot \hat{y}_i | 
+ |\sum_{i=1}^n \hat{x}_i \cdot \hat{y}_i - \hat{z} |
+ | \hat{z} - \hat{\hat{z}} | \\
& \leq &
| \sum_{i=1}^n (x_i - \hat{x}_i) \cdot y_i | 
+ | \sum_{i=1}^n \hat{x}_i \cdot (y_i - \hat{y}_i)  | 
+ |\sum_{i=1}^n \hat{x}_i \cdot \hat{y}_i - \hat{z} |
+ | \hat{z} - \hat{\hat{z}} | 
\end{eqnarray*}

The first term, which is itself a dot product, can be
bounded in various ways, such as by 
$\|x - \hat{x}\|_{1} \cdot \|y\|_{\infty} = (\sum_{i=1}^n |x_i - \hat{x}_i |) \cdot \max_{1 \leq i \leq n} |y_i|$,
$\|x - \hat{x}\|_{2} \cdot \|y\|_{2}$ (Cauchy-Schwartz), or
$\|x - \hat{x}\|_{\infty} \cdot \|y\|_{1}$; the second
term is analogous. Thus it suffices to bound the error of
converting each $x_i$ to $\hat{x}_i$ (and analogously for $y_i$ to
$\hat{y}_i$). If there is no conversion, or if the internal format
is higher precision than the input precision, then $x=\hat{x}$
(and $y=\hat{y}$). If the conversion is from floating point to
block floating point, then the error could be as large as
$| x_i - \hat{x}_i | \leq \epsilon \| x_i \|_{\infty}$ where
$\epsilon$ depends on the length of the fixed point format used
within the block floating point format. If the conversion is
from a higher to lower precision floating point format, then
$| x_i - \hat{x}_i | \leq \epsilon | x_i | + UN$, where
$\epsilon$ is the machine precision of the internal format,
and $UN$ is the error from underflow (which depends on whether
flush-to-zero or gradual underflow is used).

The third term, $|\sum_{i=1}^n \hat{x}_i \cdot \hat{y}_i - \hat{z} |$,
is the error in actually computing the dot product.
If this is done in floating point with machine precision $\epsilon$
and underflow error $UN$, the error can be bounded by
\[
|\sum_{i=1}^n \hat{x}_i \cdot \hat{y}_i - \hat{z} | \leq
f(n) \cdot \epsilon \cdot \sum_{i=1}^n |\hat{x}_i| \cdot |\hat{y}_i| + g(n) \cdot UN
\]
where $f(n)$ and $g(n)$ depend on the order of summation, ranging
from $f(n)=n$ for conventional serial summation to 
$f(n) = \lceil \log_2 n \rceil + 1$ when using a binary tree;
$g(n) \leq 2n$ in both cases (see \cite{Demmel84} for a more detailed
analysis of underflow). We note that if one knows that only $m < n$ of 
the summands $\hat{x}_i \cdot \hat{y}_i$ are nonzero, due to sparsity,
then one can replace $n$ by $m$ in these bounds.

We also note that there is a large literature
on using instructions like fused-multiply-add, and floating-point
tricks like two-sum, to compute dot products with arbitrarily higher
accuracy. These could possibly be of interest in using low-precision
floating point to simulate higher precision.

If the dot product is computed with block floating point using
a fixed point to represent each number, then the error bound depends
on the width of the fixed point accumulator used to sum the
fixed point products. If it is wide enough to compute the sum exactly,
the error is zero. If it is necessary to right-shift the integer
parts of $\hat{x}_i$ and $\hat{y}_i$ to avoid integer overflow, 
then the error can be bounded by
\[
|\sum_{i=1}^n \hat{x}_i \cdot \hat{y}_i - \hat{z} | \leq
n \cdot \epsilon \cdot \max_{1 \leq i \leq n} |\hat{x}_i| \cdot \max_{1 \leq i \leq n} |\hat{y}_i| 
\]
Other approaches to avoiding overflow, with different error bounds,
are imaginable. 

In the case of posits, where the precision $\epsilon$
depends on the magnitude
of the numbers (numbers closer to 1 are more accurate than much
larger or smaller numbers), we refer to \cite{GoodBadUgly} and \cite{Demmel87}, the latter of which discusses an older proposed
format with similar properties.
We also note that the designers of posits propose including a
"quire", a long accumulator capable of computing long dot products exactly. 

Long accumulators for
exact dot products with standard floating point
arguments have also been proposed
\cite{ExactDotProduct}.

Finally, the last error term, $|\hat{z} - \hat{\hat{z}}|$ bounds the
error when converting the final result from the internal format $\hat{z}$ to
the output format $\hat{\hat{z}}$. If the output is lower precision than the input,
then this can be bounded by $\epsilon |\hat{z}| + UN$, where 
$\epsilon$ and $UN$ are for the output format.

\subsection{Consistent exception handling from
input to output}

This refers to propagating Infs and NaNs that appear in the input data of a SparseBLAS
operation in a “consistent” way to the output. Sparsity makes the definition of “consistent”
more subtle than the one proposed for the dense BLAS in \cite{Proposed_BLAS_LAPACK_exception_handling}, which says that an Inf or NaN
that is input or created during execution should propagate to the output, unless there
is a good mathematical reason it should not. Sparsity makes this more subtle because
a sparse matrix contains implicit zeros, which are not stored and not expected to participate in
arithmetic operations, and so not create NaNs (from 0*NaN=NaN or 0*Inf=NaN) to
propagate. For example in SpMV, computing y = A*x, if x(i) = NaN and the i-th column
of A is all implicit zeros, then x(i)=NaN will not cause any entry of y to be a NaN, which
is the user’s expectation. 

But in fact a sparse matrix may contain 3 kinds of zeroes, that we may treat differently:
\begin{enumerate}
\item An implicit zero. As described above, this is an entry A(i,j) that is not stored in the 
user’s input data structure, and the user does not expect it to participate in any
arithmetic operations. We emphasize ``user’s input data structure’’ to distinguish
from case (3) below.
\item An explicit zero. This is a zero stored in a data structure, and it is
expected to participate in the same arithmetic operations that a nonzero value would.
The data structure could be the user’s input data structure, or one that is created by
the system to optimize performance (examples below).
\item An explicit masked zero. This refers to a zero that is explicitly stored in a data structure,
but another part of that data structure (a ``mask’’) indicates that it is not supposed to
participate in any arithmetic operations, like an implicit zero. 
\end{enumerate}

Here are two examples of why the system may create a new data structure with
explicit zeros replacing some implicit zeros. First, being able to access and
operate on a segment of matrix entries of some fixed length makes it easier to
use SIMD instructions on some architectures. Second, it may only be necessary
to store one column index for such a segment instead of multiple indices
(e.g. blocked CSR, or BCSR), saving memory and reducing memory access time.

One example of a user’s input data structure that can contain explicit masked zeros is ELLPACK, 
in which the nonzeros of an m x n sparse matrix are stored in an m x k matrix A , and row i of A
contains the nonzeros in row i of the sparse matrix, where k is the maximum number of 
nonzeros in any row. This means row i of A is padded with additional explicit zeros if it contains
fewer than k nonzeros. There is also an m x k Col matrix containing the column indices of the
matrix entries stored in A, with an index of -1 to mask out any padded zeros.

To be ``consistent’’ given these different kinds of zeros, we require both that
implicit zeros and explicit masked zeros do not create NaNs which would then
propagate to the output. 

So far we have discussed the SpMV operation defined as $y=A \cdot x$. 
If it were defined as $y = \alpha \cdot A \cdot x + \beta \cdot y$, as in the dense BLAS,
then we need to consider the cases $\alpha = 0$ and $\beta = 0$. In the dense BLAS,
$\beta = 0$ is defined to mean that the requested operation is $y = \alpha \cdot A \cdot x$,
i.e. the input $y$ is never read, only written, so any Infs or NaNs it may contain are
irrelevant, and should not propagate. Similarly, $\alpha = 0$ is defined to mean
$y = \beta \cdot y$, so Infs and NaNs in $A$ or $x$ are irrelevant, and never accessed.
And if $\alpha = \beta = 0$, the vector $y$ is simply initialized to 0.

We note that that we are not trying define all the cases in which an entry of $y$
must be an Inf or a NaN. To see why, suppose the result
is the sum of four numbers: $[z, z, -z, -z]$ where $z$ is a finite number greater than
half the overflow threshold. Then depending on the order of summation, the result
could be NaN, +Inf, -Inf or 0 (the correct answer). And if there is a fifth summand
equal to +Inf, the result could be +Inf or NaN. So consistency requires only that
an Inf input propagates to the output as an Inf or NaN; a NaN will naturally
propagate as a NaN. The issue of getting the bitwise identical result every time SpMV
is performed is the separate question of reproducibility, discussed below.

Finally, for the more complicated case of solving a triangular
system of equations, see \cite{Proposed_BLAS_LAPACK_exception_handling}
for a discussion of the dense case.

\subsection{Bit-wise reproducibility from run to run} \label{sec:BitwiseReproducibility}

There are times where getting exactly the same result (bitwise identical output) from multiple runs of a math routine is important, for instance in finance, scientific research, cryptography, code debugging, and many others.  However, it is also well known that in floating point arithmetic, addition and multiplication are not always associative, so that the order in which array elements are multiplied and/or added together affects the truncation and rounding and ultimately the final result, so bitwise identical results are often not guaranteed.  The concept of "conditional numerical reproducibility (CNR)" has been introduced in some performance math libraries to address this desire to have algorithms or functions with clear conditions for consistently obtaining reproducible floating point results.  Note that the choice to enable conditional numerical reproducibility may come with a sacrifice of performance. Those conditions under which reproducibility is guaranteed have converged to four common scenarios, each with more stringent level of care required in the algorithm design:
\begin{description}
\item [Default (no reproducibility property).] Some run to run variation in output results may occur based on potentially different orders of floating point arithmetic operations within the algorithm.  These are often the most performant algorithms.
\item [CNR property.] Reproducible results when calls occur within the same executable where the number of computational threads the library uses remains constant through the run. Can be less performant than the default case.
\item [Strict CNR property.] Reproducible results when run within the same executable where the number of computational threads used within the algorithm can vary. Can be less performant than CNR property algorithms.
\item [Full Numerical Reproducibility.] Reproducible results when calls occur from different executables on the same machine or on separate machines including on separate computer architectures.  Can be less performant than Strict CNR property algorithms.
\end{description}

While the "Full Numerical Reproducibility" property might seem the most desirable, it can be extremely difficult to achieve on modern architectures and would likely sacrifice a great amount of performance, so is unlikely to be something supported by any Sparse BLAS library.  We will restrict the remaining discussions to the support of the CNR and strict CNR properties in Sparse BLAS, which means run to run reproducibility within a single executable and all things equal except possibly the number of computational threads.

While it is common for some sparse formats and algorithms to be naturally reproducible due to the method of parallelizing the work (for instance using row-based parallelism for SpMV in CSR format where a single computation thread processes the multiply-adds with the dense vector across a sparse matrix row in the order that the column indices were stored in that row), there are other cases where reductions, atomic operations or general algorithms with multiple computational threads contributing to the floating point result that require more care in ensuring the order of floating point operations is maintained from run to run.  We leave it to the Sparse BLAS function implementer to come up with algorithmic ways to ensure such reproducible results can happen and will turn our attention now to how a user could signal to the function that such care is required.

We propose to introduce a global CNR property for the library that can be set via the following APIs. We also recommend that when algorithms are available for selection in an API, that library implementers denote in documentation the  algorithms which have the different CNR properties themselves.  This would allow a fine grained approach to be applied by users as well without the global switch being set. If the global property and the local algorithm selection do not align, the local algorithm selection will be preferred.

\begin{listing}[H]
\begin{minted}{c++}
namespace spblas {

    enum class cnrProperty {
        default = 0,
        none = 0,
        cnr = 1,
        strict_cnr = 2
    };

    void set_cnr_property(cnrProperty prop); 

    cnrProperty get_cnr_property();

}
\end{minted}
\caption{Conditional Numerical Reproducibility properties in Sparse BLAS}
\end{listing}

Finally, we note that this notion of conditional numerical reproducibility is  orthogonal to the notion of consistent exception handling discussed previously. Both can be set or not set independently. There is a small intersection, but   The combination of ``permissive`` and ``cnr`` would mean that you always get an Inf or a NaN in the same output location from run-to-run, but it may not be ``consistent`` in the sense that we cannot guarantee which NaN payload is propagated in an operation like NaN+NaN, just that one of the arguments is propagated (see sec. 6.2.3 of the IEEE 754 standard).

\subsection{Test Suites}
Non-associativity of floating-point arithmetic complicates the process of testing numerical code.
The most common approach is to check if a computed quantity is within some theoretically informed error bound of a reference value.
The particular error bound used and the way that the reference values are constructed varies with the test suite implementation.
Indeed, most developers of BLAS libraries maintain their own test suites.
The test suite for reference BLAS is available at www.netlib.org/lapack as part 
of the LAPACK release,
and if these tests are not satisfied by an individual test, then the test is deemed 
a failure.

For the Sparse BLAS effort, we will deploy a unit test framework that compares the accuracy of functionality implementations against a reference implementation, typically a sequential CPU code relying on the CSR matrix format. The test suite will then compose of a set of sparse matrices covering a wide range of sparsity structures and numerical properties.

\section{Future extensions}
As hinted at in Section~\ref{sec:sparse-formats}, we also recognize the value of many other sparse formats that are not covered so far in this proposal, like the CSR-4 array format, the Block CSR formats, the Ellpack format, and it's several variants and many others found in literature and existing sparse BLAS libraries.  Subsequent extensions of this proposal should address the addition of more sparse matrix formats to the sparse BLAS support.  We do recommend that format conversions between these additional formats should likely always go back through one of the original core CSR/CSC/COO sparse matrix formats to limit the scope and potential need for conversion routines.  Likewise, as the formats expand, each implementation will need to define how they will handle matrix products or matrix additions between disparate input sparse formats, and denote which format pairs for inputs are actually supported.

Functionality-wise, we plan to expand the sparse triangular solves to also accept dense matrices and sparse vectors, and extend the scope to more complex algorithms like solvers and graph algorithms.

We also recognize the increasing use of tensors, which are higher-dimensional analogs of matrices, in computing. 

Sparse tensor computations such as tensor decompositions and contractions present unique challenges ranging from sparse tensor formats and different functions such as sparse tensor times tensor multiplication and sparse matricized tensor times Khatri-Rao product (MTTKRP).

Our current design is not optimal for representing and computing on tensors, even when they are matricized, but we plan to provide extensions of our API for sparse tensor computations.

\bibliographystyle{plain}


\end{document}